\documentclass[12pt]{article}

\usepackage{secdot}

\newenvironment{hangref}
  {\begin{list}{}{\setlength{\itemsep}{4pt}
  \setlength{\parsep}{0pt}\setlength{\leftmargin}{+\parindent}
  \setlength{\itemindent}{-\parindent}}}{\end{list}}

\def\qed{\relax\ifmmode\hskip2em \fbox{ }\else\unskip\nobreak\hskip1em 
$\fbox{}$\fi}

\newsavebox{\theorembox}
\newsavebox{\lemmabox}
\newsavebox{\corollarybox}
\newsavebox{\propositionbox}
\newsavebox{\examplebox}
\newsavebox{\propertybox}
\savebox{\theorembox}{\bf Theorem}
\savebox{\lemmabox}{\bf Lemma}
\savebox{\corollarybox}{\bf Corollary}
\savebox{\propositionbox}{\bf Proposition}
\savebox{\examplebox}{\bf Example}
\savebox{\propertybox}{\bf Property}

\newtheorem{definition}{{\sc Definition}\rm }[section]
\newtheorem{definitions}[definition]{{\sc Definitions\rm }}

\newlength{\proofskip}
\begin{document}

\begin{center}

{\LARGE 
A Fast Template Based Heuristic For Global Multiple Sequence Alignment}
% \footnote{A preliminary version of this Report 2010-34}\\[12pt]

% Authors and addresses:

\footnotesize

\mbox{\large Srikrishnan Divakaran, Arpit Mithal, and Namit Jain}\\

DA-IICT, 
Gandhinagar, Gujarat, India 382007,
\mbox{
srikrishnan-divakaran@daiict.ac.in,
arpit-mithal@daiict.ac.in, and 
namit-jain@daiict.ac.in }\\[6pt]
\normalsize
\end{center}

% begin "double spacing" the text:
\baselineskip 20pt plus .3pt minus .1pt

% Here is the abstract:

\noindent 
\begin{abstract}
Advances in bio-technology have made available massive amounts of functional, structural and genomic data for many biological sequences. This increased availability of heterogeneous biological data has resulted in biological applications where a multiple sequence alignment (msa) is required for aligning similar features, where a feature is described in structural, functional or   evolutionary terms. In these applications, for a given set of sequences, depending on the feature of interest the optimal msa is likely to be different, and sequence similarity can only be used as a rough initial estimate on the accuracy of an msa. This      has motivated the growth in template based heuristics that    supplement    the     sequence information with evolutionary, structural and functional   data and exploit feature similarity instead of sequence similarity to construct multiple sequence alignments that are biologically more accurate. However, current frameworks for designing template based heuristics do not allow the user to explicitly specify information that can help to classify features into types and associate weights signifying the relative importance of a feature with respect to other features, even though in many instances this additional information is readily available. This has resulted in the use of ad hoc measures and algorithms to define feature similarity and msa construction respectively. \newline \newline
In this paper, we first provide a mechanism where as a part of the template information the user can explicitly specify for each feature, its type, and weight. The type is to classify the features into different categories based on their characteristics and the weight signifies the relative importance of a feature with respect to other features in that sequence. Second, we exploit the above information to define scoring models for pair-wise sequence alignment that assume segment conservation as opposed to single character (residue) conservation. Finally, we present a fast progressive alignment based heuristic framework that helps in constructing a global msa by first constructing an msa involving only the informative segments using exact methods, and then stitch into this the alignment of non-informative segments constructed using fast approximate methods.
\end{abstract}
\bigskip
% Here are the Keywords:
\noindent\hrulefill

% The body of the paper starts here:
%******************************************************
\section{Introduction}
A global multiple sequence alignment (msa) [7, 17, 29] of  a 
set $\cal{S}$ = $\{S_1, S_2, ..., S_k\}$ of $k$ related protein     sequences is  a  way  of arranging the characters in $\cal{S}$ into a rectangular grid of columns by introducing zero or more spaces into each sequence so that similar sequence features   occur   in   the same column, where a feature can be any  relevant biological information like secondary/tertiary structure, function, domain decomposition, or  homology to the common ancestor. The goal in attempting to construct a global msa is either to identify conserved features that may explain their functional, structural, evolutionary or phenotypic similarity, or identify mutations that may explain functional, structural, evolutionary or phenotypic variability.  \newline \newline
Until recently, sequence information was the only information that was easily available for many proteins. So, the measures that were used to evaluate the quality (accuracy) of   a msa were mostly based on sequence similarity. The {\em sum of pairs score} (SP-score) and {\em Tree score} were two such measures that were widely used. For both these measures, the   computation   of    an optimal msa is known to be NP-Complete [54]. So, in practice most of the focus is on designing fast approximation algorithms and heuristics. From the perspective of approximation algorithms, constant polynomial time approximation algorithms are known for the SP-score [17,55] and polynomial time approximation schemes (PTAS) [55] are known for the Tree score. However, in practice, these approximation algorithms have large run-times that makes them not very useful even for moderate sized problem instances. From the perspective of heuristics, most heuristics are based on progressive alignment [18, 20, 48, 49, 51], iterative alignment [10, 11, 12, 14, 15, 19, 24, 47], branch and bound [45], genetic algorithms [36, 37], simulated annealing [26] or on Hidden Markov   Modeling (HMM) [8, 9, 21]. For an extensive review of the various heuristics for msa construction, we refer the reader to excellent survey articles of Kemena and Notredame [25], Notredame [33, 34, 35], Edger and Batzoglou [12], Gotoh [15], Wallace et al. [52], Blackshields et al. [5]. \newline \newline
In heuristics based on progressive alignment, the msa is constructed by first computing pair-wise sequence distances using optimal pair-wise global alignment scores. Second, a clustering algorithm (UPGMA or NJ [46]) uses these pair-wise sequence distances to construct a rooted binary tree, usually referred to as guide tree. Finally, an agglomerative algorithm uses this guide tree to progressively align sequences a pair at a time to construct a msa. The   pair-wise global alignments scores are usually computed using a substitution matrix and   a     gap penalty scheme that is based on sequence similarity. ClustalW [51] was among the first widely used progressive alignment tool on which many of the current day progressive aligners are based. In this   paper, our focus is on heuristics that are based on progressive alignment mainly because in this method the computation of pair-wise sequence distances, guide tree and the choice of agglomerative algorithm for progressively pair-wise aligning sequences can be essentially split into three independent steps. This helps to provide a flexible algorithmic framework for designing simple parameterized greedy algorithms that are computationally scalable and whose parameters can be tuned easily to improve its accuracy. In addition, the alignments obtained through   this approach are usually a good starting point for other popular approaches like iterative, branch and bound, and HMM. However, the progressive aligners because of their greedy   approach    commit mistakes early in the alignment process that are usually very hard to correct even when using sophisticated iterative aligners. This problem can be addressed if we can incorporate into the pair-wise scoring scheme the information for every pair of sites the frequency at which the residues at these sites are involved in alignments involving other sequences in $\cal{S}$.  However, incorporating    this  information for all pairs of sites based on all sequences in $\cal{S}$  is computationally infeasible. \newline \newline
The consistency based heuristics [6, 10, 11, 24, 28, 38, 39, 40, 42, 43, 47] 
tackle this problem by incorporating a larger fraction of this information at a reasonable computational cost as follows: The score for aligning residues at a pair of sites is estimated from a collection of pair-wise residue alignments    named     the        library. The library is constituted of pair-wise alignments whose residue      alignment characteristics are implicitly assumed to be similar to an optimal msa or a reference alignment that was constructed using sequence independent methods. For a given   library, any pair     of residues receives an alignment score equal to the number of times these two residues have been found aligned either directly or indirectly through a third residue. The Consistency based progressive aligners generally construct msas that are more accurate than the pure progressive aligners like clustalW.  However, it is not very clear how to construct   a   library of alignments whose reside alignment characteristics are guaranteed to be similar to an optimal msa. In addition, the increased accuracy of msa of consistency based aligners comes at a computational cost that is on an average $k$ times more than a pure progressive aligner. T-Coffee [38], ProbCons [6], MAFFT [24], M-Coffee [53], MUMMALS [41], EXPRESSO [2], PRALINE [42], T-Lara [4] are some of the widely used consistency based aligners. \newline \newline
Currently, advances in bio-technology have made available massive amounts of functional, structural and genomic data  for many biological sequences. This increased availability of heterogeneous   biological data has resulted in biological applications where an msa is required for aligning similar   features, where   a feature is described in structural, functional or   evolutionary terms. In these applications, for a given set of sequences, depending on the feature of interest the optimal msa is likely to be different, and sequence similarity can only be used as a rough initial estimate on the accuracy of an msa. In addition, from evolutionary studies we know that structure and function of biological sequences are usually more conserved than the sequence itself. This      has motivated the growth in template based heuristics [50] that    supplement    the     sequence information with evolutionary, structural and functional   data and exploit feature similarity instead of sequence similarity to construct multiple sequence alignments that are biologically more accurate. In these methods, each sequence is associated with a template, where a template can either be a 3-D structure, a profile or prediction of any kind. Once a template is mapped onto a sequence, its information content can be used to guide the sequence alignment in a sequence independent fashion. Depending on the nature of the template one refers to its usage as structural extension or      homology extension. Structural extension takes advantage of the increasing number of sequences with an experimentally characterized homolog in the PDB database, whereas homology   extension uses profiles. 3-D Coffee [3], EXPRESSO [3], PROMALS [42, 44] and PRALINE [47] are some popular aligners that are widely used tools that employ template based methods. For more details about template based methods we refer the reader to Kemena and Notredame [25] and Notredame [34]. \newline \newline
In template based methods, we can view each template once mapped to a sequence as essentially partitioning the sequence into segments, where each segment corresponds to a feature described by the template. Then, we construct a msa by essentially aligning segments that share similar features. The current frameworks for describing templates do not allow the user to explicitly specify information that can help (i) classify features into types and (ii) associate a weight signifying the relative importance of a feature with respect to other features, even though in many instances this additional information is readily available. This has resulted in the use of ad hoc measures and algorithms to define feature similarity and msa construction respectively. \newline \newline 
In this paper, we 
\begin{itemize}
\item [-] provide a mechanism where as a part of the template information the user can explicitly specify for each feature, its type, and weight. The type is to classify the features into different categories based on their characteristics and the weight signifies the relative importance of a feature with respect to other features in that sequence. 
\item[-] define scoring models for pair-wise sequence alignment that assume segment conservation as opposed to single character (residue) conservation. Our scoring schemes for aligning pairs of segments are based on segment type, segment weight, information content of an optimal local alignment involving that segment pair, and its supporting context. This is an attempt to define scoring schemes that evaluate a pair-wise global alignment through information content of a global segment alignment, where segments correspond to features within sequences. For example, in     a structurally correct alignment the focus is on aligning residues that play a similar role in the 3D structure of the sequences, whereas a   correct alignment from an evolutionary viewpoint focuses on aligning two residues that share a similar relation to their closest     common   ancestor, and in a functionally correct alignment the focus is on aligning residues that are known to be responsible for the same function. The supporting context consists of set of sequences that are known to belong to the same family (i.e. share similar structure, function or homology to a common ancestor) as the       given sequence pair  and can help determine to what extent the alignment of the features in that pair-wise alignment is consistent with         the alignment of these features with other sequences in the family.  
\item[-] present a fast progressive alignment based heuristic   that essentially constructs global msa by first classifying segments into informative or non-informative segments based on their information content determined using segment scoring matrices. Then, using exact methods, we construct  a global msa involving only the informative segments. Finally, using approximate methods we construct the alignment of non-informative segments and stitches them into the alignment of informative segments.
\end{itemize}
{\bf Remark}: The statistical theory for evaluating alignments in terms of its information content was  developed for local alignments by Karlin and Altschul [23]. However, their theory do not extend to the case of global alignments. The pair-HMMS provide a framework for statistical analysis of pair-wise global alignments for complex scoring schemes using standard methods like Baum-Welch and Viterbi training. However, determining the right set of parameters for optimal statistical support is highly non-trivial and involves dynamic programming algorithms with computational complexity that is quadratic in the length of the given sequences. 
\newline \newline
The rest of     this    paper is structured as follows. In Section $2$, we define the problem and introduce the relevant terms and notations to define our segment scoring schemes and heuristics. In Section $3$, we present our segment scoring schemes. In Section $4$, we present our heuristics, in Section $5$, we describe our experimental set-up and summarize our preliminary experimental results, and in Section $6$,
we present our conclusions and future work.
\section{Preliminaries}
In  this section, we first define the problem of msa construction for a given a set of sequences and their segment decompositions, where each segment is classified into one of many types and is  associated       with a weight reflecting its importance relative to other segments within that sequence. Then, we introduce some basic terms and definitions that are required for defining our scoring models and heuristics.
\subsection{Problem Definition}
Let $\cal{S}$ = \{ $S_1,…, S_k$ \} be a set of $k$ related protein sequences each of length $n$. For $i \in [1..k]$,  let $B_i = \{ B_i^{1},…, B_i^{n_{i}} \}$ be the decomposition of $S_i$ into $n_i$ segments. Each segment $s \in B_i$ is classified into one of many types based on the type of features that are known/predicted to be present in that segment, and is associated with a non-negative real number weight that reflects the importance of the feature associated with that segment relative to other features in that sequence. That is, each segment $s \in B_i$, $i \in [1..k]$, is associated with a type $type(s)$, and  a non-negative real number weight $weight(s)$. 
\newline \newline
{\bf Example}: If the sequences in $\cal{S}$ are partitioned into segments based on their predicted secondary structure, each segment is classified into one of three types {\em helix}, {\em strand} or a {\em coil}, is associated with a non-negative weight in the interval $[1, 10]$ that reflects the confidence in its secondary structure classification. \newline \newline
Given a set $\cal{S}$ of $k$ biological related sequences, their decomposition into segments, and the type and weight associated with each of these segments, our goal is to design fast progressive alignment based heuristics that exploit the information content in these segments to build a biologically significant multiple sequence alignment. 
\subsection{Basic Terms and Definitions}
Now we introduce some terms and definitions that will necessary for defining our segment scoring models and heuristics.
\begin{definitions}
For $i \in [1..k]$, we define 
\begin{itemize}
\item[-] $B_i^{inf}=\{s \in B_i: weight(s) \geq \alpha \}$ to be the segments in $S_i$ whose weights are greater  than or equal     to $\alpha$, where $\alpha$ is a non-negative user specified real number parameter. We refer to 
the segments in $B_i^{inf}$ as {\em informative segments} of $S_i$;
\item[-] $S_i^{inf}$ to be the subsequence of $S_i$ obtained by concatenating the segments in $B_i^{inf}$ in the order in which they appear in $S_i$. We refer to this subsequence as the {\em informative sequence} of $S_i$.
\end{itemize}
\end{definitions}
\begin{definition}
For a pair of segments $s \in B_i^{inf}$ and $t \in B_j^{inf}$ of the same type, $i\ne j \in [1..k]$, we define
$L^{H}(s,t)$ to be the local alignment between $s$ and $t$ constructed using heuristic $H$ and BLOSUM62 
scoring matrix and $SEG^{H}(s,t)$ to be bit score correpsonding to $L^{H}(s,t)$. 
\end{definition}
\begin{definitions}
For $i \ne j \in [1..k]$, we define
\begin{itemize}
\item[-] $\alpha_{i,j}$ to be a real number in the interval [0,2] that reflects the level of divergence between $S_i$ and $S_j$.
We estimate the level of divergence between $S_i$ and $S_j$ using the bit score of a local alignment between $S_i^{inf}$ and 
$S_j^{inf}$ constructed using heuristic $H$ and $BLOSUM62$ scoring matrix;
\item[-] $c:[0-2] \rightarrow R^{+}$ is a function that computes for a given level of divergence the information threshold for an 
alignment to be informative. 
\end{itemize}
\end{definitions}
\begin{definitions}
For a segment $s \in B_i^{inf}$ and $j \ne i \in [1..k]$, we define
\begin{itemize}
\item[-] $Neighbor_{j}(s) = \{ t \in B_{j}^{inf} : type(t) = type(s) \land SEG^{H}(s, t) \geq c(\alpha_{i,j})*|s|\}$ to be the set of 
informative segments $t$ in $S_j$ of the same type as $s$ with bit score of a local alignment between $s$ and $t$ greater than or 
equal to $c(\alpha_{i,j})*|s|$. We refer to the segments in $Neighbor_{j}(s)$ to be the neighbors of $s$ in $S_j$.
\item[-] $Closest-neighbor_{j}(s) = \{ u' \in B_j^{inf} : SEG^{H}(s, u') =  max_{t \in Neighbor_{j}(s)}^{} SEG^{H}(s, t)\}$  is   
the neighbor of $s$ in $S_j$ that maximizes  the bit score of a pair-wise local alignment with $s$. 
We refer to such a segment to be the closest neighbor of $s$ in $S_j$.
\item[-] $Neighborhood(s) = \bigcup_{ j \in [1..k]}^{} Closest-neighbor_{j}(s)$ to be the set    consisting of a closest neighbor 
of $s$ from each sequence in $\cal{S}$.
\end{itemize}
\end{definitions}
\begin{definitions}
For $i \in [1..k]$,
\begin{itemize}
\item[-] $B_i^{nei} =\{ s \in B_i$ : $s$ is a neighbor of some segment in $\cal{S}$ $\setminus S_i$. \}. 
\item[-] $S_i^{nei}$ to denote the subsequence obtained by concatenating the  segments in $B_i^{nei}$ in the order in which they appear in $S_i$. We refer to this subsequence as the {\em neighbor sequence} of $S_i$.
\end{itemize}
\end{definitions}
\begin{definitions}
For each pair of segments $s \in S_i^{nei}$ and $t \in S_j^{nei}$ of the same type from distinct sequences in $\cal{S}$, and  $l \ne i,j \in [1 ..k]$, we define
\begin{itemize}
\item[-] $Mutual-neighbors_{l}(s,t) =\{u \in B_{l}^{nei} : u \in Neighbor_{i}(u)  \bigcap  Neighbor_{j}(u) \}$ to be the segments in $S_l$ that are neighbors of both $s$ and $t$.
\item[-] $Closest-mutual-neighbor_{l}(s, t) = \{ u' \in B_l^{nei} : 
SEG^{H}(s, u') + SEG^{H}(t, u') = max_{u \in mutual-neighbors_{l}(s, t)}^{} (SEG^{H}(s, u) + SEG^{H}(t, u))$        
to  be  the  mutual  neighbor $u$ of $s$  and $t$  in $S_l$ that maximizes $SEG^{H}(s, u)+ SEG^{H}(t, u)$. We 
refer to such a segment as the closest mutual neighbor of $s$ and $t$ in $S_l$.
\item[-] $Mutual-neighborhood(s,t) = \bigcup_{ j \in [1..k]}^{} Closest-mutual-neighbor_{j}(s,t)$   to  be 
the set consisting of a closest mutual neighbor of $s$ and $t$ from each sequence in $\cal{S}$.
\end{itemize}
\end{definitions}
\section{Scoring Models for Global Segment Alignment}
In    this   section, we define scoring models for pair-wise segment alignment of sequences.  
We           classify        segments     into informative and non-informative based on their weight    and construct   segment scoring        matrix entries only for informative segments. 
Restricting the segment scoring matrix entries to only informative segments helps to significantly reduce the computational time 
of our heuristics with minimal impact on alignment accuracy. In Section $3.1$, we introduce scoring schemes for aligning pairs of 
informative segments. In Section $3.2$, we introduce scoring schemes for aligning a segment with a gap.
\subsection{ Scoring Schemes for Aligning Pairs of Segments}
 We now introduce   the following three scoring schemes for aligning a pair $s,t$ of informative segments of the same type: 
(i) Progressive scoring; (ii) Linear Consistency scoring, and (iii) Quadratic Consistency scoring. \newline \newline
{\bf Progressive scoring}: $SCORE(s, t) = SEG^{H}(s,t)$. In this scheme, we only make use of the information content of a pair wise local alignment between segments $s$ and $t$ constructed using heuristic $H$ and BLOSUM62 scoring matrix. \newline \newline
{\bf Linear consistency scoring}: \newline
$SCORE(s, t)$ = $|Mutual-neighborhood(s,t)|$ * $\sum_{u \in Mutual-neighborhood(s, t)}^{} 
(SEG^{H}(s, u) + SEG^{H}(t, u))$. In this scheme, we make use of the information from both (i) pair-wise local alignment between $s$ and $t$, and (ii) pair-wise alignments involving the segments $s$ and $t$ with segments in their mutual neighborhood. \newline \newline 
%This scheme incorporates consistency based information and attempts to %balance %he higher bit score obtained through few pair-wise alignments 
%between not very highly conserved segments from closely related %sequences with %he information obtained from many pair-wise alignments 
%of conserved segments from diverging sequences. \newline \newline
{\bf Quadratic consistency scoring}:  \newline
$SCORE(s, t) = |Mutual-neighborhood(s, t)|^{2}$ $* \sum_{u \in Mutual-neighborhood(s, t)}^{}$
$(SEG^{H}(s, u) + SEG^{H}(t, u))$. In this scheme, the information obtained through the alignment of two conserved segments of two diverging sequences is weighed more than the information obtained through the alignment of two non-conserved segments of two closely related sequences. 
%In this scheme, we define 
%In instances where the average divergence between pairs of sequences in %$\cal{S}$ is high, 
%we define the quadratic consistency model where 
%the information obtained through the alignment of two conserved segments %of %two diverging sequences is weighed more than the 
%information obtained through the alignment of two non-conserved segments %of %two closely related sequences. In this scheme, we define 
\subsection{ Scoring Schemes for Aligning a Segment with a Gap}
We now introduce the following two scoring schemes for aligning an informative segment with a gap: (i) Zero gap penalty, and (ii) Maximum gap penalty. \newline \newline
{\bf Zero gap penalty}: $SCORE(s, -) = 0$.  In this scheme, we do not penalize the deletion of any informative segment. 
%This model is employed in instances where there are many segment repeats %where %the maximum gap penalty and linear gap penalty schemes %turn out %to be %ineffective and some times counter productive. 
\newline  \newline
{\bf Maximum gap penalty}: $SCORE(s, -) = \max_{t \in Neighborhood(s)}^{} SEG^{H}(s, t)$.  
In this scheme, 
%In the absence of frequent repeats this scheme is employed. In this %scheme, we %define That is, we calculate 
the gap penalty of $s \in B_i$ based on the informative segment 
$t \in \cal{S} \setminus S_i$ of the same type that maximizes the bit score of a pair-wise local alignment between $s$ and $t$. 
%In this scheme the consistency based information is not incorporated %into the %scoring. 
%\newline \newline 
%{\bf Linear gap penalty}: $SCORE(s, -) = \sum_{t \in %Neighborhood(s)}^{} SEG^{H}(s, t)$. 
%In this scheme we %incorporate consistency based information %btained through bit scores %of local alignment involving 
%$s$ with all the %nformative segments in %its neighborhood. 
\section{Heuristics for msa Construction}
We now present a generic framework for designing  template based fast progressive alignment heuristics   that construct global msa 
as follows: 
\begin{itemize}
\item[(i)] construct $DIST^{nei}$, a matrix of pair-wise sequence distances, based on scores of pair-wise global segment 
alignment involving only the informative segments; 
\item[(ii)] construct a guide tree $G^{nei}$ using $NJ$ algorithm using $DIST^{nei}$ and 
build $MSA^{nei}$, a msa of the informative segments, by progressively pair-wise segment aligning sequences consistent with 
$G^{nei}$;
\item[(iii)] construct the pair-wise global alignment of the residues in non-informative segments using fast   
approximate methods and stitch them back into $MSA^{nei}$. 
\end{itemize}
In Section $4.1$, we describe our heuristic and in Section $4.2$, we 
present our Heuristic $A$. 
\subsection{Description of Our Heuristic}
{\bf Construction of pair-wise sequence distances}: We now describe how we compute the pair-wise sequence distances for each pair of sequences in $\cal{S}$.
\begin{definitions}
For $i,j \in [1..k]$, we define  
\begin{itemize}
\item[-] {\em a global segment alignment} between two sequences $S_i$ and $S_j$ to be an alignment where a segment in $S_i$ is either 
aligned to a gap or another segment in $S_j$ of the same type;
\item[-] $G^{nei}(i,j)$ to be the optimal global segment alignment between $S_i^{nei}$ and $S_j^{nei}$ constructed using the segment 
scoring matrix $SCORE$;.
\item[-] $DIST^{nei}(i, j)$ to be the score corresponding to $G^{nei}(i,j)$.
\end{itemize}
\end{definitions}
Notice that if each segment consists of a single amino acid then the global segment alignment is the same as a traditional global alignment. In this case, the traditional scoring matrices can be used to score alignments between segments. Otherwise, one needs to determine an appropriate 
segment scoring matrix and then using Needleman-Wunsch's [32] dynamic program construct an optimal global pair-wise segment alignment. 
\newline \newline 
{\bf Construction of Guide Tree and msa of Informative Segments}:
We construct the guide tree $G^{nei}$ from pair-wise sequence distance matrix $DIST^{nei}$ using the Neighbor Joining (NJ) algorithm. Then, 
we construct $M^{nei}$, the msa of informative of sequences in $\cal{S}$, by progressively pair-wise globally segment aligning the sequences 
$S_1^{nei}, S_2^{nei}, Ã¢â‚¬Â¦, S_k^{nei}$ consistent with $G^{nei}$. \newline \newline 
{\bf Stitching the sites in non-informative segments into msa of informative sequences}:
We will describe now for each pair of sequences $S_i$ and $S_j$ that were progressively aligned while constructing $M^{nei}$, 
how we stitch the alignment of sites in $S_i$ and $S_j$ that were either in non-informative segments or in non-aligned 
portion of informative segments into $M^{nei}$.  First, we introduce some necessary definitions.
\begin{definitions}
For a pair of informative segments $s \in B_i^{inf}$ and $t \in B_j^{inf}$ of the same type, let $L^{H}(s,t)$ be the local alignment 
of $s$ and $t$ constructed using heuristic $H$ and BLOSUM62 matrix. we now define
\begin{itemize}
\item[-] $PREFIX_{s}(L^{H}(s, t))$ to be  the prefix of segment $s$ that is not part of the local alignment $L^{H}(s, t)$;
\item[-] $SUFFIX_{s}(L^{H}(s, t))$ to be the suffix of segment $s$ that is not part of the local alignment $L^{H}(s, t)$.
\end{itemize}
\end{definitions}
Let $G^{nei}(i,j), i \ne j \in [1..k]$, denote an optimal global segment alignment between sequence $S_i^{nei}$ and $S_j^{nei}$ 
constructed using the segment scoring matrix $SCORE$. 
For $G^{nei}(i,j)$, we say a segment $s \in S_i$ to be a {\em matched segment} 
if in $G^{nei}(i,j)$ it is aligned to a segment $t \in S_j^{nei}$, otherwise it is an {\em unmatched} segment.  
We now present a procedure $stitch$ that stitches the 
alignment between the sites in $S_i$ that occur between any two consecutive matched segments $s$ and $\hat{s}$ and the sites in 
$S_j$ that occur between the corresponding matched segments $t$ and $\hat{t}$ into $G^{nei}(i,j)$.
\newline \newline
{\bf Procedure $Stitch(s, \hat{s})$}: 
\begin{itemize}
\item[-] Let $p, \hat{p}$ ($q, \hat{q}$) be the respective indices of segments $s, \hat{s}$ ($t, \hat{t}$) 
in $S_i$      and $S_j$; 
\item[-] Let $A=SUFFIX_{i}(B_i^{p}, B_j^{q}) \bigcup_{l=p+1}^{\hat{p}-1} B_i^{l}$ $\bigcup PREFIX_{i}(B_i^{\hat{p}}, 
B_j^{\hat{q}})$ be the    sequence        of sites in $S_i$ that are 
either in non-informative segments or unaligned 
portions of informative segments in $G^{nei}(i,j)$; 
\item[-] $B=SUFFIX_{j}(B_i^{p}, B_j^{q}) 
\bigcup_{l=q+1}^{\hat{q}-1} B_j^{l}$ 
$\bigcup PREFIX_{j}(B_j^{\hat{p}}, B_j^{\hat{q}})$  be the    sequence of sites in $S_j$ that are 
either in non-informative segments or unaligned 
portions of informative segments in $G^{nei}(i,j)$; 
\item[-] Globally align segments $A$ and $B$ using BLOSUM62 scoring matrix and any fast linear time 
heuristic and then insert this alignment between segments $s$ and $\hat{s}$ in $G^{nei}(i,j)$.
\end{itemize}
\subsection{Heuristic $A(\alpha, H, c)$}
\begin{itemize}
\item[] Parameters:
\begin{itemize}
\item[(1)] $\alpha$: a non-negative real number;
\item[(2)] $H$: an algorithm/heuristic for pair-wise local alignment of sequences;
\item[(3)] $c$: a function that maps for any given level of divergence in the interval 
$[0,2]$ to the information threshold for an alignment to be informative. 
\end{itemize}
\item[] Inputs: 
\begin{itemize}
\item[(1)] $S = \{ S_1, ..., S_k \}$: the set of k input sequences;
\item[(2)] $B = \{B_1, ..., B_k \}$: the set consisting of the segment decompositions  of the sequences in $\cal{S}$, where each segment 
$s$ is associated with a type $type(s)$ and weight $weight(s)$;
\end{itemize}
\item[] Main Heuristic 
\begin{itemize}
\item[(1)] For $i \in [1..k]$, construct $B_i^{inf}$ = \{$s \in B_i$ : $weight(s) \ge \alpha$ \} and $S_i^{inf}$ the sequence of informative segments.
\item[(2)] For each pair of informative segments $s \in B_i^{inf}$ and $t  \in B_j^{inf}$ of the same type, using heuristic $H$ and BLOSUM62 scoring matrix construct $L^{H}(s,t)$ and compute $SEG^{H}(s,t)$.
\item[(3)] For $i \ne j \in [1..k]$, set $\alpha_{i,j}$ to the bit score per unit length corresponding to $L^{H}(S_i^{inf}, 
S_j^{inf})$, the local alignment between $S_i^{inf}$ and $S_j^{inf}$ constructed using heuristic $H$ and 
BLOSUM62 scoring matrix.
\item[(4)] For each informative segment $s \in B_i^{inf}$ and $j \ne i \in [1..k]$, compute the following: 
\begin{itemize}
\item[(i)] $Neighbor_{j}(s) = \{ t \in B_{j}^{inf} : type(t) = type(s) \land SEG^{H}(s, t) \geq c(\alpha_{i,j})*|s|\}$.
\item[(ii)] $Closest-neighbor_{j}(s) = \{ u' \in B_j^{inf} : SEG^{H}(s, u') =  max_{t \in Neighbor_{j}(s)}^{} SEG^{H}(s, t) \}$.
\item[(iii)] $Neighborhood(s) = \bigcup_{j \in [1..k]}^{} Closest-neighbor_{j}(s)$.
\end{itemize}
\item[(5)] For each sequence $S_i \in \cal{S}$, $i \in [1..k]$, construct 
$B_i^{nei} =\{ s \in B_i$ : $s$ is a neighbor of some segment in $\cal{S}$ $\setminus S_i$. \} and $S_i^{inf}$, the neighbor sequence of $S_i$.

\item[(6)] For each pair of segments $s \in S_i^{nei}$ and $t \in S_j^{nei}$ from distinct sequences and  $l \ne i,j 
\in [1 ..k]$, compute
\begin{itemize}
\item[(i)] $Mutual-neighbors_{l}(s,t) =\{u \in B_{l}^{nei} : u \in Neighbor_{i}(u)  \bigcap  Neighbor_{j}(u) \}$.
\item[(ii)] $Closest-mutual-neighbor_{l}(s, t) = \{ u' \in B_l^{nei} : SEG^{H}(s,u') + SEG^{H}(t, u') =
max_{u \in mutual-neighbors_{l}(s, t)}^{} (SEG^{H}(s, u) + SEG^{H}(t, u)) \}$.
\item[(iii)] $Mutual-neighborhood(s,t) = \bigcup_{ j \in [1..k]}^{} Closest-mutual-neighbor_{j}(s,t)$.
\end{itemize}
\item[(7)] For each segment $s \in B_i^{nei}$, $i \in [1..k]$,  compute $SCORE(s,-)$.
\item[(8)] For    each    pair of segments $s \in B_i^{nei}$ and $t \in B_j^{nei}$ of the same type, compute  $SCORE(s, t)$.
\item[(9)] For  $i \ne j \in [1..k]$, compute $DIST^{nei}(i,j)$ by 
globally segment aligning $S_i^{nei}$ and $S_j^{nei}$ using Needleman-Wunch's dynamic program and 
segment scoring matrix $SCORE$.
\item[(10)] We now construct the msa of $\cal{S}$ as follows:
\begin{itemize}
\item[(i)] Construct guide tree $T^{nei}$ from  $DIST^{nei}$ using  the Neighbor Joining (NJ) algorithm.
\item[(ii)] Construct $M^{nei}$ by progressively globally segment aligning the sequences $S_1^{nei},  ...,   S_k^{nei}$ 
a pair at a time consistent with $T^{nei}$.
\item[(iii)] For each pair of sequences $S_i^{nei}$ and $S_j^{nei}$ that       were progressively aligned while constructing $M^{nei}$, 
\begin{itemize}
\item[-] Let $G^{nei}(i, j)$ denote the global segment alignment of $S_i^{nei}$ and $S_j^{nei}$;
\item[-] For each pair $s, \hat{s}$ of consecutive matched segments of $S_i^{nei}$ in $G^{nei}(i,j)$ 
(where $t, \hat{t}$ are the corresponding matched segments of $S_j^{nei}$)
use {\em procedure stitch} to stitch the alignment between the sites in $S_i$ that occur between $s$ and 
$\hat{s}$ and the sites in $S_j$ that occur between the sites in $t$ and $\hat{t}$ to $G^{nei}(i, j)$. 
\end{itemize}
\end{itemize} 
\end{itemize}
\end{itemize}
\section{Experimental Results}
In this section, we first describe our experimental set-up, then we describe how we evaluate the performance of our 
heuristic, and finally  we summarize our preliminary experimental results. 
\subsection{Experimental Set-up}
Our computational 
experiments have been set-up with the focus on analyzing the performance of our heuristics for 
sequences from protein families in the PFAM [13] database for which (i) accurate reference alignments were available 
either through structural aligners or through other sequence independent biological methods, and (ii) annotations describing 
the salient biological features were available for each sequence. We chose 12 sets of sequences ranging from 5 to 23 sequences with sequence similarity ranging from 20\% to 80\%. For these sequences, we used PSIPRED [22], a widely used structure prediction tool, to partition 
each sequence into segments based on their secondary structure characteristics. PSIPRED classifies each segment into one of three 
types {\em helix}, {\em strand} or a {\em coil}, and associated a non-negative weight in the interval $[1, 10]$ reflecting the 
confidence in its partitioning and classification. 
Then for these sets of sequences, we construct an msa by using our heuristic $A(\alpha, H, c)$, 
where $\alpha$ is a non-negative real 
number parameter for classifying segments based on their weights into 
informative and non-informative segments, $H$ is an algorithm/heuristic for pair-wise local alignment of segments, and $c$ is a function 
that maps for any given level of divergence in the interval $[0,2]$ to the information threshold for an alignment to be informative.  
In our experiments, we have set $\alpha$ to be $6$. That is a segment is considered to be informative if its average segment weight $\geq 6$ (i.e. $\alpha \geq 6$) and its length is at least 5. In addition, if two informative segments    of the same type are separated by less than 4 residues we merged the two segments with the intervening residues into a single informative segment. We set $H$, the algorithm/heuristic for local alignment to be BLASTP [1, 30] 
with slight modifications to handle alignments involving short sequences. \footnote{The quality of alignment constructed using Smith-Waterman’s dynamic program was not significantly different from that obtained using BLASTP.} We defined the function $c$ based on the average bit scores of BLOSUM matrices corresponding to different levels of sequence divergence. 
\subsection{Evaluation the Performance of Our Heuristic}
We evaluate the performance of our heuristic based on
(i) the accuracy of its msa in comparison with an reference alignment, and (ii) its computational efficiency for the appropriate choice 
of its parameters $\alpha$, $H$ and $c$, \newline \newline 
{\bf Evaluating accuracy of an msa}: The traditional sequence similarity based measures like SP score and Tree score have only been helpful in  providing a crude estimate of the alignment quality and measures based on structurally correct alignments are likely to be better alternatives for evaluating alignment accuracy. So, for sequences for which their 3D structure is known, the accuracy of an msa 
can be evaluated in comparison with reference alignments constructed through a structure aligner. 
We also observe instances of homologous sequences that share only a few features and 
yet preserve their overall structure and function. In these instances, local feature conservation is another good predictor of 
alignment accuracy. So, we measure the accuracy of the msas constructed by our heuristic in terms of the percentage correlation 
between the columns in the multiple sequence alignments constructed    by our heuristic and the columns of the sites within the 
reference alignment that correspond to conserved features. \newline \newline
{\bf Note}:
Our heuristics make use of the secondary structure predictions from PSIPRED. So, any inaccuracies in the secondary structure prediction of PSIPRED should also be factored while evaluating the    accuracy of msa constructed by our heuristics. We factor this in terms of the correlation between the informative sites in our heuristic and the sites in the reference alignment that correspond to conserved features. We also restrict the impact of inaccuracies in secondary structure prediction on msa accuracy by conservative choice of the information threshold function $c$ (i.e. higher than if we had an accurate partitioning and correct classification of segments). \newline \newline
{\bf Evaluation of Computational Efficiency}: 
Our heuristics attempt to minimize its computational time with minimal impact on its accuracy by first 
classifying the segments within each sequence into informative (non-informative) segments based on its 
weight exceeding (not exceeding) $\alpha$. Then, the msa is essentially constructed by first progressively 
pair-wise aligning the sites in informative segments using exact methods and then us linear time 
approximate heuristics to align the sites in non-informative segments and stitch them back into the 
alignment of sites in informative segments. So, the     saving in computational time depends on the fraction      
 of the segments that are informative. This in turn depends mainly on the choice of the information threshold 
$\alpha$. 
\subsection{Summary of Preliminary Experimental Results}
\begin{table}[ht]
%\caption{Summary of msa results using Linear Consistency and Max Gap Penalty Schemes} % title of Table
\centering % used for centering table
\begin{tabular}{c c c c c c c} % centered columns (4 columns)
\hline\hline %inserts double horizontal lines
Protein & \# of & Sequence & 
\% Sequence &  \# of informative &
Avg Length of & 
\% Local \\ 
Family & Sequences & Lengths & 
Similarity &  segments &
informative segment & 
Similarity \\ 
%Case & Method\#1 & Method\#2 & Method\#3 \\ [0.5ex] % inserts table
%heading
\hline % inserts single horizontal line
PF13420 & 21 & 152-164 & 20\%-70\% & 4 & 10 & $<70\%^{a}$ \\ 
PF13652 & 11 & 131-152 & 65\%-85\% & 4 & 12 & $>90$\% \\ 
PF13693 & 22 & 77-81   & 55\%-83\% & 3 & 12 & $>90$\% \\ 
PF13733 & 5  & 133-142 & 55\%-61\% & 2 & 8 & $>90$\% \\ 
PF13844 & 6  & 449-481 & 68\%-78\% & 7 & 12 & $>90$\% \\ 
PF13856 & 23 &  90-112 & 30\%-73\% & 3 & 10 & $>80\%^{a}$ \\ 
PF13944 & 21 & 120-146 & 30\%-85\% & 3 & 10 & $>90$\% \\ 
PF14186 & 11 & 152-157 & 38\%-68\% & 4 & 8 & $>90$\% \\ 
PF14263 & 10 & 120-129 & 50\%-66\% & 3 & 10 & $>90$\% \\ 
PF14274 & 20 & 155-165 & 36\%-71\% & 3 & 12 & $>90$\% \\ 
PF14323 & 18 & 485-548 & 36\%-43\% & 6 & 11 & $>90$\% \\ 
% inserting body of the table
%1 & 50 & 837 & 970 \\ % inserting body of the table
[1ex] % [1ex] adds vertical space
\hline %inserts single line
\end{tabular}
$a$: Quadratic Consistency and Max Gap Penalty Scoring Schemes was employed. \\ 
\label{table:nonlin} % is used to refer this table in the text
\caption{Summary of msa results using Linear Consistency and Max Gap Penalty Schemes} % title of Table 
\end{table}
%\footnote{For this test case Quadratic Consistency and Max Gap Penalty was employed}\% \\ 
\section{Conclusions and Future Work}
Our preliminary experimental results indicate that our template based heuristic framework can help in 
designing heuristics that can exploit template based information to construct msas that are 
biologically accurate in a computationally efficient manner. However, we would like to 
(i) make use of {\em extreme value distribution} [16] to define the the function $c$ that maps 
for a given level of sequence divergence the information threshold for an alignment to be informative;
(ii) Understand how to define the segment scoring schemes for  aligning sequences that are highly 
divergent; (iii) evaluate the accuracy of the alignments constructed by our heuristics by using  
sequence independent measures [2, 25, 34] on challenging datasets in BAliBASE [27, 28].
\section*{References}
\begin{hangref}
\item{[1]}
Altschul SF, Gish W, Miller W, Myers EW and Lipman DJ.
Basic local alignment search tool. J. Mol. Biol. 1990;215:403-410. 
\item{[2]} Armougom F, Moretti S, Keduas V, Notredame C., 
The iRMSD: a local measure of sequence alignment accuracy using structural information. 
Bioinformatics. 2006 Jul 15;22(14):e35-e39.
\item{[3]} 
Armougom F, Moretti S, Poirot O, Audic S, Dumas P, Schaeli B, Keduas V and Notredame C. 
Expresso: automatic incorporation of structural information in multiple sequence alignments using 3D-Coffee. 
Nucleic Acids Res, 2007;Jul 1-34.
\item{[4]}
Bauer M, Klau GW, and Reinert K. 
Accurate multiple sequence-structure alignment of RNA sequences using combinatorial optimization. 
BMC Bioinformatics, 2007;8(271).
\item{[5]}
Blackshields G, Wallace IM, Larkin M, and Higgins DG.
Analysis and comparison of benchmarks for multiple sequence alignment. 
In Silico Biol. 2006;6(4):321-39.
\item{[6]}
Do CB, Mahabhashyam MS, Brudno M and Batzoglou S.
ProbCons: Probabilistic consistency-based multiple sequence alignment.
Genome Res. 2005 Feb;15(2):330-40.
\item{[7]} 
Carrillo H and Lipman DJ.
The Multiple Sequence Alignment Problem in 
Biology. 
SIAM Journal of Applied Mathematics.
1988;Vol.48, No. 5, 1073-1082.
\item{[8]} 
Durbin R, Eddy S, Krogh A and Mitchison G. Biological sequence 
analysis: probabilistic models of proteins and nucleic acids, Cambridge 
University Press. 1998.
\item{[9]}
Eddy SR. 
Multiple alignment using hidden Markov models. 
Proc of Int Conf Intell Syst Mol Biol (ISMB). 
1995;3:114-20.
\item{[10]}
Edgar RC. 
MUSCLE: a multiple sequence alignment method with reduced time and space complexity. 
BMC Bioinformatics. 2004a;5:113. 
\item{[11]}
Edgar RC. 
MUSCLE: multiple sequence alignment with high accuracy and high throughput.
Nucleic Acids Res. 2004b;32:1792–1797. 
\item{[12]}
Edgar RC and Batzoglou S. 
Multiple sequence alignment. 
Curr. Opin. Struct. Biol. 2006;16:368–373.
\item{[13]}
Finn RD, Mistry J, Tate J, Coggill P, Heger A, Pollington JE, 
Gavin OL, Gunesekaran P, Ceric G, Forslund K, Holm L, Sonnhammer EL, Eddy SR 
and , Bateman A.
The Pfam protein families database.
Nucleic Acids Res. 2010;Database Issue 38:D211-222
\item{[14]} 
Gotoh O. 
Consistency of optimal sequence alignments. 
Bull. Math. Biol. 1990;52:509–525. 
\item{[15]} 
Gotoh O. 
Significant improvement in accuracy of multiple protein sequence alignments by iterative 
refinements as assessed by reference to structural alignments. 
J. Mol. Biol. 1996;264:823–838. 
\item{[16]}
Gumbel EJ. Statistics of extremes. Columbia University Press, New York, NY. 1958.
\item{[17]} 
Gusfield D.
Algorithms on Strings, Trees and Sequences: Computer 
Science and Computational Biology. 
Cambridge University Press. 1997.
\item{[18]} 
Higgins DG and Sharp PM. 
CLUSTAL: a package for performing multiple 
sequence alignment on a microcomputer. 
Gene 1998;73(1): 237–244. 
\item{[19]} 
Hirosawa M, Totoki Y, Hoshida M and Ishikawa M. Comprehensive 
study on iterative algorithms of multiple sequence alignment. 
Comput Appl Biosci 1995;11 (1): 13–18. 
\item{[20]}
Hogeweg P and Hesper B. 
The alignment of sets of sequences and the construction of phylogenetic trees.  An integrated method. 
J. Mol. Evol. 1984;20:175–186. 
\item{[21]} 
Hughey R and Krogh A. Hidden Markov models for sequence analysis: 
extension and analysis of the basic method. CABIOS 1996;12 (2): 95–107. 
\item{[22]}
Jones DT. 
Protein secondary structure prediction based on 
position-specific scoring matrices. J. Mol. Biol. 
1991;292: 195-202. 
\item{[23]}
Karlin Samuel and Altschul Stephen F. (1990). 
Methods for assessing the statistical significance of 
molecular sequence features by using general 
scoring schemes. 
Proc Natl Acad Sci USA 1987 (6): 2264–8. 
\item{[24]}
Katoh K and Toh H. 
Recent developments in the MAFFT multiple sequence alignment program. Brief. 
Bioinform. 2008;9:286–298.
\item{[25]}
Kemena C and Notredame C. 
Upcoming challenges for multiple sequence alignment methods in the high-throughput era.
Bioinformatics. 2009;Oct 1;25(19):2455-65. 
\item{[26]} 
Kim J, Pramanik S and Chung MJ. 
Multiple sequence alignment using 
simulated annealing. 
Comput Appl Biosci 1994;10 (4): 419–26. 
\item{[27]} 
Lassmann T and Sonnhammer EL. 
Quality assessment of multiple alignment programs. 
FEBS Lett. 2002;18:126–130. 
\item{[28]}
Lassmann T and Sonnhammer EL. 
Automatic assessment of alignment quality. Nucleic Acids Res. 
2005a;33:7120–7128. 
\item{[29]} 
 Lipman DJ, Altschul SF and Kececioglu JD. 
A tool for multiple 
sequence alignment. 
Proc Natl Acad Sci U S A 1988:86 (12): 4412–4415. 
\item{[30]}
McWilliam H, Valentin F, Goujon M, Li W, Narayanasamy M, Martin J, Miyar T and Lopez R. 
Web services at the European Bioinformatics Institute - 2009
Nucleic Acids Res. 2009;37: W6-W10.
\item{[31]}
Morgenstern B. 
Multiple DNA and Protein sequence based on segment-to-segment comparison. 
Proc. Natl Acad. Sci. USA. 1996;93:12098–12103. 
\item{[32]} 
Needleman SB and Wunsch CD. 
A general method applicable to the search for similarities in the amino acid sequence of two proteins. 
J. Mol. Biol. 1970;48:443–453. 
\item{[33]}
Notredame C. 
Recent evolutions of multiple sequence alignment. PLoS Comput. Biol. 2007;3:e123. 
\item{[34]}
Notredame C. 
Computing Multiple Sequence Alignment with Template-Based Methods. 
Sequence Alignment: Methods, Models and Strategies, Edited by
Michael S. Rosenberg, University of California Press. 
2011;Chapter 4:56-68.
\item{[35]}
Notredame C. 
Recent evolutions of multiple sequence alignment. PLoS Comput. Biol. 2007;3:e123. 
\item{[36]} 
Notredame C, O'Brien EA and Higgins DG. 
SAGA: RNA sequence alignment 
by genetic algorithm. 
Nucleic Acids Res. 1997 25 (22): 4570–80. 
\item{[37]}
Notredame C and Higgins DG. 
SAGA: sequence alignment by genetic algorithm. Nucleic Acids Res. 1996;24:1515–1524. 
\item{[38]}
Notredame C, Higgins DG and Heringa J.
T-Coffee: A novel method for fast and accurate multiple sequence alignment.
J Mol Biol. 2000 Sep 8;302(1):205-17.
\item{[39]}
O'Sullivan O, Suhre K, Abergel C, Higgins DG and Notredame C.
3DCoffee: combining protein sequences and structures within multiple 
sequence alignments. J. Mol. Biol. 2004;340:385–395.
\item{[40]}
Pei J. 
Multiple protein sequence alignment. 
Curr. Opin. Struct. Biol. 2008;18:382–386. 
\item{[41]}
Pei J and Grishin NV. 
MUMMALS: multiple sequence alignment improved by using hidden Markov models with local structural information. 
Nucleic Acids Res. 2006;34:4364–4374. 
\item{[42]}
Pei J and Grishin NV. 
PROMALS: towards accurate multiple sequence alignments of distantly related proteins. 
Bioinformatics. 2007;23:802–808. 
\item{[43]} 
Pei J, Sadreyev R and Grishin NV.
PCMA: fast and accurate multiple sequence alignment based on profile 
consistency. Bioinformatics. 2003;19:427–428. 
\item{[44]}
Pei J, Kim BH and Grishin NV.
PROMALS3D: a tool for multiple protein sequence and structure alignments. 
Nucleic Acids Res. 2008;36:2295–2300. 
\item{[45]}
Reinert K, Lenhof H,  MutzelP, Mehlhorn K and Kececioglou JD.
A branch-and-cut algorithm for multiple sequence alignment.
Recomb97. 1997;241-249.
\item{[46]}
Saitou N and Nei M. 
The neighbor-joining method: a new method for reconstructing phylogenetic trees. 
Molecular Biology and Evolution, volume 4, issue 4, pp. 406-425, July 
\item{[47]}
Simossis VA and Heringa J. 
PRALINE: a multiple sequence alignment toolbox that integrates homology-extended and secondary structure information. 
Nucleic Acids Res. 
\item{[48]}
Subramanian AR, Weyer-Menkhoff J, Kaufmann M and Morgenstern B.
DIALIGN-T: an improved algorithm for segment-based multiple sequence alignment.
BMC Bioinformatics. 2005 Mar 22;6:66.
\item{[49]}
Subramanian AR, Kaufmann M and Morgenstern B.
DIALIGN-TX: greedy and progressive approaches for segment-based multiple sequence alignment.
Algorithms Mol Biol. 2008 May 27;3:6. doi: 10.1186/1748-7188-3-6.
\item{[50]}
Taylor WR. Identification of protein sequence homology by consensus template alignment. 
J. Mol. Biol. 1986;188:233–258. 
\item{[51]}
Thompson JD, Higgins DG and Gibson TJ.
CLUSTAL W: improving the sensitivity of progressive multiple sequence alignment 
through sequence weighting, position-specific gap penalties and weight matrix choice.
Nucleic Acids Res. 1994 Nov 11;22(22):4673-80.
\item{[52]}
Wallace IM, O'Sullivan O and Higgins DG.
Evaluation of iterative alignment algorithms for multiple alignment. 
Bioinformatics. 
2005b;21:1408–1414. 
\item{[53]}
Wallace IM, 'Sullivan O, Higgins DG and Notredame C.
M-Coffee: combining multiple sequence alignment methods with T-Coffee. 
Nucleic Acids Res. 2006;34:1692–1699. 
\item{[54]} 
Wang L and Jiang T. On the complexity of multiple sequence alignment. 
J Comput Biol. 1994;1 (4): 337–348. 
\item{[55]}
Wang L, Jiang T and Gusfield D.
A more Efficient Approximation Scheme for Tree Alignment.
SIAM Journal on Computing. 1997;Vol 30, No. 1, 283-299.
%\item{[56]}
%Zhou H and Zhou Y.
%SPEM: improving multiple sequence alignment with sequence profiles and predicted secondary structures.
%Bioinformatics. 2005 Sep 15;21(18):3615-21. Epub 2005 Jul 14.
\end{hangref}
\end{document}